\documentclass{elsart}

 \usepackage{epsfig}

\usepackage{amssymb}

\begin{document}

\begin{frontmatter}



\title{Three-boson recombination at ultralow temperatures}


\author{M. T. Yamashita}
\ead{yamashita@itapeva.unesp.br}
\address{Universidade Estadual Paulista, 18409-010, Itapeva, SP, Brazil}
\author{T. Frederico}
\ead{tobias@ita.br}
\address{Departamento de F\'\i sica, Instituto Tecnol\'ogico
de Aeron\'autica, 12228-900, S\~ao Jos\'e dos Campos, SP, Brazil}
\author{Lauro Tomio}
\ead{tomio@ift.unesp.br}
\address{Instituto de F\'\i sica Te\'orica, Universidade
Estadual Paulista, 01405-900, S\~{a}o Paulo, Brazil}

\begin{abstract}
The effects of trimer continuum resonances are considered in the
three-body recombination rate of a Bose system at finite energies
for large and negative two-body scattering lengths ($a$). The
thermal average of the rate allows to apply our formula to Bose
gases at ultra-low temperatures. We found a good quantitative
description of the experimental three-body recombination length of
cesium atoms to deeply bound molecules up to 500 nK. Consistent
with the experimental data, the increase of the temperature moves
the resonance peak of the three-body recombination rate to lower
values of $|a|$ exhibiting a saturation behavior.
\end{abstract}

\end{frontmatter}

% main text

Efimov states~\cite{EfPLB70} result from a general phenomenon that
occurs in three-boson systems, corresponding to the appearance of
an increasing number of bound states as the two-body binding
energy tends to zero. The number is infinite when the two-body
energy is zero. So, for systems close to an infinite two-body
scattering length some curious three-boson dynamical effects are
expected to appear due to the existence of weakly bound trimers or
continuum resonances. Presently the universal properties of
three-boson systems can be exploited in its plenitude in ultracold
atomic traps by varying from positive to negative values the
two-body scattering length ($a$) near a Feshbach resonance.

Recent experimental observation of a signature of Efimov physics
was reported in Ref.~\cite{grimm06}, considering an ultracold gas
of cesium atoms in a trap. Such results definitely open a window
to study the universal behavior of three-body systems with large
scattering lengths in a controlled way, and can be considered as
an experimental verification of correlations among few-body
observables with short range interactions with respect to the
scattering length.

A giant three-body recombination loss for a large and negative
two-body scattering length, where the three-body state hits the
continuum threshold~\cite{BrPRL01}, was reported in
Ref.~\cite{grimm06}.  For a particular value of the scattering
length, characteristic of their experiment, the three-body state
attains zero binding energy and produces a huge increase in the
measured recombination rate when three free-atoms with low kinetic
energy collide forming deeply bound dimers. The increase of the
gas temperature $T$ from 10 nK to 250 nK depletes the recombination
peak moving it to lower values of $|a|$. This behavior indicates
that the zero energy bound state evolves to a triatomic continuum
resonance~\cite{ressonancia}, in agreement with results obtained
for the trajectory of  an Efimov state in the complex plane, i.e.,
the zero energy trimer dives into the continuum forming a
resonance when $|a|$ is increased~\cite{ressonancia}.

Theoretically, the position and width of the triatomic resonance
moves to larger values as $|a|$ diminishes, starting from the one
that allows a zero energy state~\cite{ressonancia}. This agrees
with the observation that as the temperature raises the peak 
moves to smaller $|a|$. The average energy available for the
recombination process raises with temperature, consequently the
recombination peak (roughly at the resonance position) moves
toward smaller values of $|a|$ and the intensity diminishes due
to the increase of the decay rate to the continuum states. The
resonance decay to the three-body continuum states competes with
the formation of the deeply bound dimer.

The theory for calculating the zero temperature recombination
rate~\cite{BrPRL01,braaten06} represents fairly well the 10 nK
data. The issue now is how to describe quantitatively the
temperature dependence of the recombination rate and the peak
dislocation toward smaller $|a|$ by raising $T$.

In the present work, we generalize the formula of Braaten and
Hammer for the recombination rate into deeply bound states. 
The expression, derived for the  particular case of zero
kinetic energies~\cite{BrPRL01,braaten06} is extended to positive
energies. We include the effect of continuum trimer resonances
for $a<0$ in the recombination rate. A thermal average of the
recombination rate is performed before quantitative comparison
with the experimental results of Kraemer et al.~\cite{grimm06} for
the cesium gas at $T=$ 10, 200 and 250 nK. The key point to
generalize the rate for positive energies, is to write down the
energy and width of continuum resonances in terms of $a<0$ and a
three-body scale. To obtain the detailed form of these functions,
we solved numerically the renormalized zero-range three-body
model~\cite{ressonancia,AdPRL95,virtual} in the complex plane to
compute the continuum resonance energy and width as a function of
$a$ and a three-body scale.

In practice it is needed one three-body information to set the
scale of a real system. For example, in the experiment of
Ref.~\cite{grimm06} performed with a 10 nK trapped cesium atoms
near a Feshbach resonance there is an evidence of the virtual
formation of a zero energy state, during the three-atom
recombination process to a deeply bound dimer. The recombination
peak was found at the particular value of $a$ in their fitting
procedure. This information is enough to set the three-body scale.

The three-body physical scale is given by the binding energy ($B_3$) of the
shallowest trimer just below the zero-energy Efimov state. The
trimer binding energy $B_3$ is calculated from the universal
relation~\cite{ressonancia}
\begin{eqnarray}
a_-^{-1}=0.0297\sqrt{\frac{m}{\hbar^2}B_3} ~, \label{a}
\end{eqnarray}
where $\hbar^2/m=48.12/m$~K\AA$^2$, and $m$ is the boson mass number. 
In the following general discussion, the three-body scale will be given
by the particular value of the scattering length $-a_-$, where the
three-boson system has an excited Efimov state at zero energy.

The recombination rate, for a finite energy $E$, can be written as
\begin{eqnarray}
L_3(E)\; = \;3 C(a,E)\frac{\hbar a^4}{m}, \label{rate3}
\end{eqnarray}
where $C(a,E)$ at zero energy is given by~\cite{braaten06}:
\begin{equation}
C(a,0)=4590\frac{\sinh(2\eta_-)}{\sin^2[s_0\ln(|a|/a_-)]+\sinh^2\eta_-},
\label{crate3}\end{equation}
where $\eta_-$ is a dimensionless quantity that characterizes the 
resonance width. It describes the unknown decay rate of Efimov states 
into deeply bound dimer states plus a free atom. 

Equations (\ref{rate3}), with $L_3\equiv L_3(0)$, and (\ref{crate3}) 
were used in Ref.~\cite{grimm06} 
to fit the recombination length $\rho_3$ at zero energy, as 
\begin{equation}
\rho_3=\left[\frac{2m}{\sqrt{3}\hbar} L_3\right]^{\frac14}.\label{rho}
\end{equation}
The finite energy extension of Eq.~(\ref{crate3}) is done by the
inclusion of the pole brought by the continuum resonance. 

With the trimer at zero energy, the S-matrix pole is obtained from
\begin{eqnarray}
\zeta(a,a_-)=\sin[s_0\ln(|a|/a_-)],
\label{zeta0}\end{eqnarray}
when $a=-a_-$ ($\zeta(a,a_-)=0$). Here and in the following, $a_-\equiv a_-(E=0)$ 
to simplify the notation. In a naive extension of the above function 
to the complex energy plane, the corresponding zeros,
$\zeta(a,a_-(E))=0$, will give the energy and partial width of the 
continuum resonances
related to the scattering length $a=-a_-(E)$.
 
In extending Eq.~(\ref{crate3}) to non-zero energies, it is assumed that 
the parameter $\eta_-$, which characterizes the width of the resonance, 
varies slowly with the scattering energy. 
This implies that it is not expected an appreciable effect when the trimer 
moves from the threshold to an ultralow energy continuum resonance.
However, the calculated decay width of the resonances to continuum 
states has a magnitude large enough to be important for the observed results.

The slow variation of $\eta_-$ in terms of the energy is expected because it 
corresponds to the formation of a deeply bound dimer with a size much smaller 
than $|a|$. Therefore, the short-range part of the triatomic wave function 
should be responsible for the recombination process, which is deeply inside 
the potential, such that does not strongly depend on the wave function tail,
determined by the large scattering length.

The problem now is focused on how to get $a_-$ as a
function of energy. Or, how to obtain $a_-(E)$.
The resonance energy and partial width scale with $a$ and $a_-$
($a_-$ gives the dependence on the three-body scale). Therefore,
we can write that
\begin{eqnarray}
E= \frac{\hbar^2}{m~a_-^2}{\cal F}\left({a}/{a_-}\right)
~,\label{erpar1}
\end{eqnarray}
where the scaling function ${\cal F}({a}/{a_-})$ should be
universal in the limit of large values of $a$ and $a_-$,  with
respect to the interaction range. Note that the imaginary part of
$E$ in Eq.(\ref{erpar1}) gives the decay width to the continuum
states. For $a=-a_-$ the trimer has zero energy implying that
${\cal F}(-1)=0$. By expanding the real and imaginary parts 
of ${\cal F}(z)$ around $z=-1$, 
we obtain the approximate formula
\begin{eqnarray}
\frac{ma_-^2}{\hbar^2}\;E\equiv \epsilon = \alpha \left(1+\frac{a}{a_-}\right)
-{\rm i} \gamma \left(1+\frac{a}{a_-}\right)^2 ~,\label{erpar}
\end{eqnarray}
where $\alpha$ and $\gamma$ are fitting parameters.
This expression is found to reach reasonable results when comparing to the 
numerical results of the model presented in Ref.~\cite{ressonancia}.

\begin{figure}[thb]
\centerline{\epsfig{figure=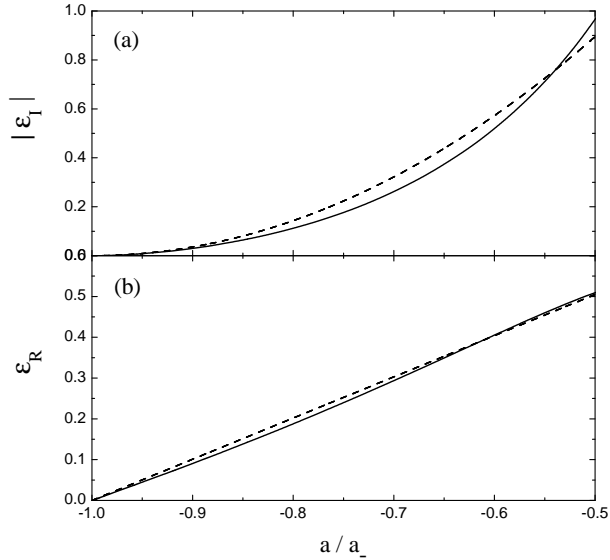,width=8.0cm}}
\caption[dummy0]{Real, $\epsilon_R$ (lower frame (b)), and
imaginary, $\epsilon_I$ (upper frame (a)), parts of the
dimensionless expression for $\epsilon \equiv
E{(ma_-^2/\hbar^2)}$ as a function of $(a/a_-)$. 
The solid lines are obtained from the renormalized zero 
model~\cite{ressonancia} and the dashed lines
are the corresponding analytic results given by Eq.~(\ref{erpar}) 
with $\alpha=1.01$ and $\gamma=3.58$.}\label{fig1}
\end{figure}

In figure~\ref{fig1} we show the dimensionless products
$\epsilon_R=Re(\epsilon)$ and $\epsilon_I= Im(\epsilon)$ as
functions of $a/a_-$, for the three-body energy scale derived from
the condition of an Efimov state at zero energy for $a=-a_-$. The
simple parameterization is compared to the calculated values using
a renormalized zero-range three-boson model~\cite{ressonancia},
with the three-body scale adjusted to allow an Efimov state at
zero energy. The real part of the resonance energy behaves almost
linearly for the range of values of $a/a_-$ shown in the figure,
while the imaginary part of $\epsilon$ has an almost quadratic
dependence on $a/a_-$.  
The values of the dimensionless constants
of Eq.~(\ref{erpar}), $\alpha=1.01$ and $\gamma=3.58$, are 
chosen to fit the numerical results obtained with the
renormalized zero range model of Ref.~\cite{ressonancia}.

%%%%%%%%%%%%%%%%%%%%%%%%%%%%%% FIG. 1 %%%%%%%%%%%%%%%%%%%%%%%%%%%%%
%%%%%%%%%%%%%%%%%%%%%%%%%%%%%%%%%%%%%%%%%%%%%%%%%%%%%%%%%%%%%%%%%%

One can easily find the function $a_-(E)$ by solving Eq.(\ref{erpar})
for a given value of $E$, with $a=-a_-(E)$:
\begin{eqnarray}
a_-(E)=a_-\left[1+{\rm i}\frac{\alpha}{2\gamma}\left(1-\sqrt{1-4{\rm i}
\frac{\gamma}{\alpha^2}\frac{ma_-^2}{\hbar^2}E}\right)\right].
\label{am}
\end{eqnarray}
Equation (\ref{am}) is strictly valid for $E$ being the resonance
energy with the imaginary part related to the partial width to the
continuum states, in spite of that we perform a naive analytic
continuation of Eq. (\ref{am}) to values of $E$ along the real
axis. Then, instead of (\ref{zeta0}), we will have
\begin{eqnarray}
\zeta(a,a_-(E))=\sin\left[s_0\ln\left(\frac{|a|}{a_- +\Delta(E)}\right)
\right] ~,\label{analy}
\end{eqnarray}
with
\begin{eqnarray}
\Delta(E)={\rm i}~a_-\frac{\alpha}{2\gamma}\left(1-\sqrt{1-4
{\rm i}\frac{\gamma}{\alpha^2}\frac{ma_-^2}{\hbar^2}E}\right) \ .
\end{eqnarray}
The function $\zeta(a,a_-(E))$ has a zero at the resonance energy
for $a=-a_-(E)$ when the decay width to the deep states vanishes.
This extends the Efimov law of appearance of bound states to the
case of continuum resonances as a function of the scattering
length. Scaling $a$ by $\exp{(\pi/s_0)}$, and considering that
$\alpha$ and $\gamma$ are scale invariants, the zero of
$\zeta(a_-(E),a)$ gives a resonance energy re-scaled by a factor
of $\exp{(-2\pi/s_0)}$.

The recombination rate for finite energy is derived from the
original expression at $E=0$ by introduction of the complex
function for $\zeta(a_-(E),a)$, given by Eq. (\ref{analy}),
assuming that the main effect for the three-body recombination
into deep dimer states comes from the change of the nearest
S-matrix pole due to the triatomic continuum resonances, while the
$\eta_-$ is a slowly varying function of $E$. The resulting
analytic expression for $L_3(E)$ in Eq.~(\ref{rate3}) is given by
\begin{eqnarray}
C(a,E)=4590\frac{\sinh(2\eta_-)}{|\zeta(a_-(E),a)+i\sinh\eta_-|^2}.
\label{rate5}
\end{eqnarray}

The next step is to perform a thermal average of the recombination rate
$\langle L_3\rangle_T$. Considering that the bosons are not condensed,
we will use a classical thermal energy distribution, which allows the separation of the
center of mass motion. The remaining degrees of freedom in the
center of mass kinetic energy $E$ are the Jacobi energies of the
two independent relative motions, then $E=E_p+E_q$, where the
index $p$ and $q$ are related, respectively, to the relative
momentum of a pair of atoms, $\vec p$, and the relative momentum of
the third atom to the pair, $\vec q$. The thermal average of
the recombination rate calculated classically, with $\beta\equiv1/(kT)$ and $k$ the Boltzmann constant, is given by
\begin{eqnarray}
\langle L_3\rangle_T= \frac{4}{\pi}\beta^3\int_0^\infty
dE_p\sqrt{E_p}\int_0^\infty dE_q\sqrt{E_q} 
\;L_3(E_p+E_q)\; e^{-[\beta(E_p+E_q)]}~.\label{rate6}
\end{eqnarray}

The three-body observables are correlated to the two- and
three-body reference scales in the Thomas-Efimov limit, where the
ratio $a/r_0$ ($r_0$ is the effective range) goes to infinity
(scaling limit). These correlations are generally defined in terms
of functions of dimensionless quantities named scaling functions.
Using $a_-$ as a reference, an observable can be written in terms
of a scaling function~\cite{virtual}, such that
\begin{eqnarray} 
{\cal{O}}\left(E \right)=(a_-)^\nu {\cal F
}\left(\epsilon,a/a_-\right)  \ , \label{o1}
\end{eqnarray}
where $\cal O$ is a general observable of the three-body system at
energy $E$ with the power $\nu$ giving the proper dimension to
the observable. The function ${\cal F }\left(\epsilon,a/a_-\right)$ converges to a
limiting curve when the Thomas-Efimov condition is
approached~\cite{virtual}. The scaling functions, calculated in
the the scaling limit~\cite{FrPRA99}, are identified with a limit
cycle~\cite{mohr}.

From the above Eq. (\ref{o1}), the thermal average of the scaling function of a given observable is
\begin{eqnarray}
\langle{\cal{O}}\left(E \right)\rangle_T=a_-^\nu \langle {\cal F
}\left(\epsilon,a/a_- \right)\rangle_T  \equiv (a_-)^\nu{\cal F
}_{th}\left(\frac{ma_-^2}{\hbar^2} kT,a/a_-\right) \ , \label{o1t}
\end{eqnarray}
which defines a correlation between $a$ and $T$ for a given value of
$\langle{\cal{O}}\rangle_T$ with the reference scale $a_-$. The
observed peak position of the recombination rate as a function of
$T$ is one of such example.

%%%%%%%%%%%%%%%%%%%%%%%%%%%%%% FIG. 2 %%%%%%%%%%%%%%%%%%%%%%%%%%%%%
\begin{figure}[thb]
\centerline{\epsfig{figure=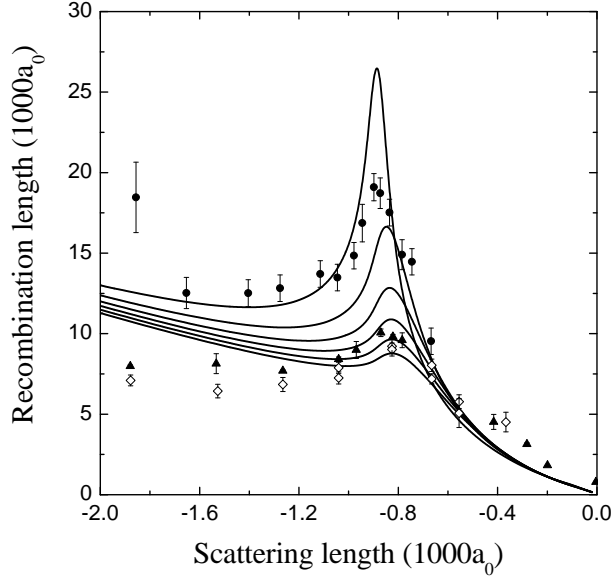,width=8.0cm}}
\caption[dummy0]{Recombination length 
$\langle\rho_3\rangle_T$ [as given by Eq.~(\ref{rhoT})]
in a trapped gas of cesium atoms, as a function of the scattering length and temperature. 
The solid curves from up to bottom are the
theoretical results for $T=$ 10 nK, 100 nK, 200 nK, 300 nK, 400 nK
and 500 nK. The symbols are the experimental results for $T=$ 10 nK
(full circles), 200 nK (full triangles) and 250 nK (open diamonds)
from \cite{grimm06}. } \label{fig2}
\end{figure}

%%%%%%%%%%%%%%%%%%%%%%%%%%%%%% FIG. 3 %%%%%%%%%%%%%%%%%%%%%%%%%%%%%
\begin{figure}[thb]
\centerline{\epsfig{figure=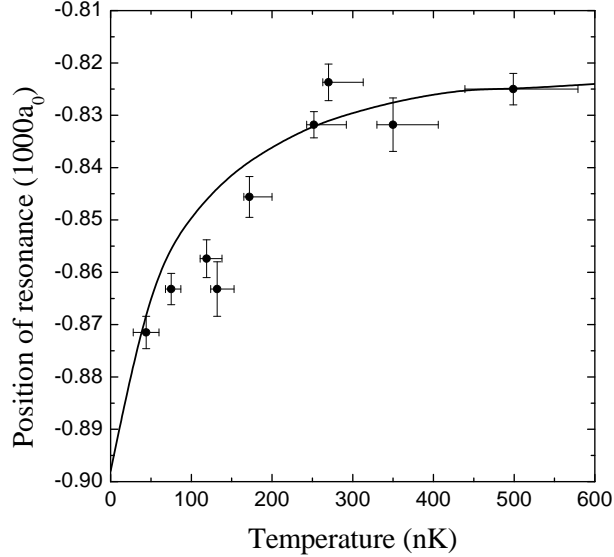,width=8.0cm}}
\caption[dummy0]{Position of the maximum of the recombination
length as a function of the temperature. For the experimental 
data, see comments in \cite{GrimmPC}.} \label{fig3}
\end{figure}
%%%%%%%%%%%%%%%%%%%%%%%%%%%%%%%%%%%%%%%%%%%%%%%%%%%%%%%%%%%%%%%%%%

The parameters of the simple curves for the resonance energy
enters through Eq. (\ref{rate5}) in the calculation of the thermal 
average of the recombination length, 
\begin{equation}
\langle\rho_3\rangle_T=\left[\frac{2m}{\sqrt{3}\hbar}\langle L_3\rangle_T
\right]^\frac{1}{4}.\label{rhoT}
\end{equation}

In figure 2, the results of the present model are compared to the experimental data given in \cite{grimm06}. 
We allowed a change in the $\eta_-$ parameter, from 0.06 found
in \cite{grimm06} to 0.03,
as we have also considered $a_-$ to be $0.898a_0$ instead of $0.85a_0$.
Replacing $a_-$ and the cesium mass number $m=133$ in Eq. (\ref{a}) we found
$B_3=1.81$ mK for the experiment. The experimental data has
temperatures around 10 nK for one set (full circles), 200 nK
(full triangles) and  250 nK (open diamonds). Our result for 200 nK is
somewhat above the data, while for 300 nK approaches the
experimental values. The increase of temperature makes the
resonance peak lower and wider, almost disappearing for 500 nK. The
movement of the peak to lower values of $|a|$ is also seen in the
figure. The present simplified model of introducing temperature dependence 
in the three-body recombination rate is also compatible with results 
obtained in Ref.~\cite{jonsell}.

In figure 3, we show the results for the scattering length
corresponding to the peak position as a function of temperature
compared to recent and preliminary experimental results~\cite{GrimmPC}. 
In the present model, no new parameters are used to obtain these results.

In summary, for a three-boson system, it was derived an extension
to positive energies of an expression given in~\cite{BrPRL01} for
the recombination length, including the triatomic continuum
resonance pole. The position and width of the triatomic resonance
increase as the modulus of the scattering length is decreased near
the critical condition for which an Efimov state turns into a
continuum resonance.

The thermal average of the observables defines new scaling
functions appropriate to ultracold gases. Our approach was tested
with experimental data of Ref.~\cite{grimm06}, as shown in Figs.
\ref{fig2} and \ref{fig3}. We found that for the particular
thermal average of the recombination rate  into deep dimer states,
the gross features of the experimental data for $T$ below 500 nK
are reproduced, with parameters set at $T=0$. The scattering
length for the peak position as a function of temperature
describes nicely the data and the saturation for $T$ above 200 nK
is well described.

To conclude, the trajectory of the triatomic continuum resonance
is reflected in the correlation curve of $a$ and $T$, which is a
direct consequence of dominance of just two scales: the two and
three body ones which determine the three-body observables.
Therefore, the thermal scaling law was checked by the experiment
with cesium atoms which strongly supports the dominance of few
scales in the physics of ultracold atoms.  Such universal thermal
scaling law for three-body recombination length into deep dimers
are expected to be confirmed also in other systems.

In view of the nice  results obtained in comparison with cesium
experimental data, this simplified analytical approach of
introducing temperature dependence in few-body thermal observables
for ultracold atomic gases is shown to give the basic properties
of a more detailed formalism to be developed.

We thank Dr. R. Grimm and T. Kraemer for useful discussion and for 
providing us preliminary data referred in \cite{GrimmPC}.
This work was partially supported by Funda\c c\~ao de Amparo a Pesquisa do 
Estado de S\~ao Paulo and Conselho Nacional de Desenvolvimento 
Cient\'\i fico e Tecnol\'ogico. 

% \appendix

% \section{}
% \label{}

\end{document}